\documentclass[aps,pre,subeqns.floatfix,amsmath,twocolumn]{revtex4}
\usepackage[pdftex]{graphicx}
\usepackage{amsmath}
\usepackage{amsfonts}
\usepackage{graphicx}
\usepackage{hyperref}
\usepackage{comment}
\usepackage{mathtools}
\usepackage{verbatim}
\usepackage[outdir=./]{epstopdf}
\usepackage{subfigure}
\usepackage{hyperref}
\usepackage{latexsym}
\usepackage{dcolumn}
\usepackage{epsf}
\usepackage{float}

\DeclarePairedDelimiter\abs{\lvert}{\rvert}

\usepackage{color}
\usepackage[utf8]{inputenc}
\begin{document}
\title{Transition to synchronization in adaptive Sakaguchi-Kuramoto model with higher-order interactions}
\author{Sangita Dutta$^1$}
\email{sangitaduttaprl@gmail.com}
\author{Prosenjit Kundu$^2$}
\author{Pitambar Khanra$^3$}
\author{Chittaranjan Hens$^4$}
\author{Pinaki Pal$^1$}
\email{ppal.maths@nitdgp.ac.in}
\affiliation{$^1$Department of Mathematics, National Institute of Technology, Durgapur~713209, India}
\affiliation{$^2$Complex Systems Group, Dhirubhai Ambani Institute of Information and Communication Technology, Gandhinagar~382007, India}
\affiliation{$^3$Department of Microbiology and Immunology, Jacobs School of Medicine \& Biomedical Sciences, The State University of New York at Buffalo, Buffalo, NY 14203, USA}
\affiliation{$^4$Center for Computational Natural Science and Bioinformatics, International Institute of Informational Technology, Hyderabad, 500032, India}

\begin{abstract}
We investigate the phenomenon of transition to synchronization in Sakaguchi-Kuramoto model in the presence of higher-order interactions and global order parameter adaptation. The investigation is done by performing extensive numerical simulations and low dimensional modeling of the system. Numerical simulations of the full system show both continuous (second order) as well as discontinuous transitions. The discontinuous transitions can either be associated with explosive (first order) or with tiered synchronization states depending on the choice of parameters. To develop an in depth understanding of the transition scenario in the parameter space we derive a reduced order model (ROM) using the Ott-Antonsen ansatz, the results of which closely matches with that of the numerical simulations of the full system. The simplicity and analytical accessibility of the ROM helps to conveniently unfold the transition scenario in the system having complex dependence on the parameters. Simultaneous analysis of the full system and the ROM clearly identifies the regions of the parameter space exhibiting different types of transitions. It is observed that the second order continuous transition is connected with a supercritical pitchfork bifurcation (PB) of the ROM. On the other hand, the discontinuous teired transition is associated with multiple saddle-node (SN) bifurcations along with a supercritical PB and the  first order explosive transition involves a subcritical PB alongside a SN bifurcation.       
\end{abstract}

\maketitle

\section{Introduction}
Network science is the keystone to study interconnected systems ~\cite{Cohen_complex_book,barabasi2012network} surrounding us such as ecosystems, social systems~\cite{shahal2020synchronization,morales2017global}, different kinds of biological~\cite{bick2020understanding,bassett2006small} and physical systems~\cite{newman2003structure,boccaletti2006complex,estrada2012structure} etc. Mathematically a network is described by a graph, where the units of a system is treated as node and the interconnection is presented by the links/edges between these nodes. If the dynamics of the isolated $i$-th node is governed by the ordinary differential equation $\dot x_i =f(x_i),~x_i\in \mathbb{R}^M$, then in the presence of pairwise interactions, the collective dynamics of the complex network is governed by the coupled equations 
\begin{equation}
    \dot x_i = f(x_i)+ K_1\sum_{j_1=1}^N A_{ij_1} G_1(x_i,x_{j_1}),~(i=1,2,\dots,N),
    \label{eq.1}
\end{equation}
where $K_1$ is the  strength of the coupling, $A = (A_{ij_1})_{N\times N}$ is the adjacency matrix of the complex network of size $N$ and $G_1(x_i,x_{j_1})$ is the coupling function between the nodes $i$ and $j_1$ of the network.  
Kuramoto introduced periodic diffusive like (sinusoidal) coupling~\cite{kuramoto1984chemical} ($G_1(x_i,x_{j_1})=\sin(x_{j_1}-x_i)$), where $f(x_i)$ depends on the intrinsic frequencies of the units. This model has been proven to successfully describe the emergent dynamics of various coupled systems. One of such interesting phenomena is synchronization~\cite{Strogatz_synchronization_book,Pikovsky_synchronization_book}. It is observed in many natural as well as man made systems like flashing of fireflies~\cite{buck1988synchronous}, clapping in a hall~\cite{neda2000sound}, brain neuron~\cite{penn2016network,cumin2007generalising}, cellular processes in populations of yeast~\cite{de2007dynamical}, metronomes, power-grids~\cite{motter2013spontaneous}, etc.
The route to synchronization from asynchronous state may be continuous, discontinuous or explosive depending on structural and dynamical variability considered in the system~\cite{gomez2011explosive,gomez2007paths,rodrigues2016kuramoto,ichinomiya2004frequency}
Moreover, the dependence of the coupling strength on the synchronization order parameter allows the model to remain updated about the dynamical state of the oscillators at each time step. This dependence can be mathematically modeled by multiplying the coupling constant with some function of order parameter. There are numerous examples of such systems that adapt the dynamic states of the units~\cite{ha2016synchronization,filatrella2007generalized}.
 
 Sakaguchi-Kuramoto (SK) model is  dynamical system which is often used to study the synchronization behavior in coupled dynamical units of phase frustrated systems, coupled with pairwise interaction\cite{sakaguchi1986soluble,kundu2017transition,kundu2019synchronization,khanra2018explosive}. The coupling function in this case takes the form $G_1(x_i,x_{j_1})=\sin(x_{j_1}-x_i-\alpha)$, where $\alpha$ is the phase frustration or phase-lag parameter. The classic scenario of SK model reveals non-hysteric continuous synchronization transition \cite{brede2016frustration}. By contrast,  an adaptive coupling (coupling constant $K_1$ is multiplied with synchronization order parameter)  may result in hysteresic or explosive transition~\cite{khanra2020amplification,khanra_chaossoli2021}. 
Here we investigate the role of higher-order interactions in SK-oscillators where the temporal synchronization order parameter will  control  the coupling strength in each time.
\par A variety of systems, such as neuronal network~\cite{petri2014homological,reimann2017cliques,sizemore2018cliques}, social network~\cite{alvarez2021evolutionary,andjelkovic2015hidden}, ecological network~\cite{billick1994higher,bairey2016high} chemical networks involve group or community interactions which can not be neglected and also have a great impact on the dynamics~\cite{musciotto2021detecting,mayfield2017higher}. For instance, the outcome of a chemical reaction between two elements can be significantly altered when a third element is introduced into the reaction. Also in case of disease infection, one healthy unit can be infected in touch of multiple infected units, takes the form of higher-order interactions~\cite{ghosh2023dimension,zhang2022epidemic}. One can also take the example of a collaboration network, where the dynamics of multi-author collaboration can not be suffices to describe with the combination of pairwise collaborations ~\cite{vasilyeva2021multilayer}. In such systems, the multi-author interactions can be represented as higher-order interactions which can not always be expressed as a sum of pairwise interactions. As a result, the concept of hypergraph~\cite{adhikari2023synchronization}, simplicial complex~\cite{jonsson2008simplicial,battiston2021physics,zhao2021simplicial,salnikov2018simplicial,iacopini2019simplicial} are introduced to encode these higher-order interactions framework. To examine the rich dynamics of a network, the higher-order interactions should be considered along with the pairwise interactions. The dynamical system with higher-order interactions can be expressed in the form,
\begin{eqnarray}
\dot x_i &=& f(x_i)+ K_1\sum_{j_1=1}^N A_{ij_1} G_1(x_i,x_{j_1}) \nonumber \\
&+& K_2\sum_{j_1=1}^N\sum_{j_2=1}^N B_{ij_1j_2} G_2(x_i,x_{j_1},x_{j_2})+\dots  \nonumber \\
&+& K_m\sum_{j_1=1}^N \dots \sum_{j_m=1}^N C_{ij_1\dots j_m} G_m(x_i,x_{j_1},\dots,x_{j_m}), \nonumber \\ 
&&\hspace{4cm} i=1,2,\dots,N  \nonumber
\end{eqnarray}
where $B_{ij_1j_2}$ and $C_{ij_1\dots j_m}$ account respectively the connections between three and $(m+1)$ units. $G_1~(K_1)$ presents the coupling function (strength) for pairwise interaction, $G_2~(K_2)$ is for triadic interaction and $G_m~(K_m)$ is for $(m+1)$ units interaction and so on~\cite{lucas2020multiorder}. The structural properties of these higher-order interactions and the impact of those structures in the emergent dynamics of networked systems draw the attention of many researchers~\cite{battiston2021physics,landry2020effect}. One of the most interesting effect of such structures is to produce explosive synchronization irrespective of the choice of natural frequencies or without any correlation between the structural and dynamical properties in different kinds of networks~\cite{skardal2019abrupt,skardal2020higher,skardal2021higher,dutta2023perfect,dutta2023impact}. \par Recently, distinct transition pathways known as tiered synchronization have been reported in the presence of both pairwise and higher-order interactions~\cite{skardal2022tiered,rajwani2023tiered}, in addition to the continuous and explosive transition to synchronization. In tiered synchronization the system transits from the incoherent state to a weak synchronization state via continuous path and then it abruptly jump from the weak synchronization state to strong synchronization state. Where as in the backward transition, depending on the hysteresis width, it jumps from the synchronization state to the weak synchronization or incoherent state.  Several modification in the Kuramoto system can induce tiered paths. For example introduction of time delay in the coupling function in a system with higher-order interactions originate tiered paths~\cite{skardal2022tiered}. Also adapting the global order parameter in the higher-order terms can give rise to tiered route to synchronization~\cite{rajwani2023tiered}. An interplay between adaptation in higher-order term in absence of any adaptation in pairwise term leads the system to transit from the explosive  to continuous paths of synchronization via tiered paths. Whereas, in absence of higher-order interactions the adaptation parameter in the pairwise term promotes only explosiveness of synchronization transitions in a phase frustrated system~\cite{khanra2020amplification}. Till now the effect of the adaptation of order parameter  in a phase frustrated system with higher-order interactions has not been explored. In this paper, we focus on the reciprocity between the global order parameter adaptation, higher-order coupling strength and the phase-lag term. Since we are aware of the role of these parameters separately in the synchronization transitions from the previous literature, we are eager to elaborate the combined effect of these parameters in a phase oscillator system. 
\par Here, we accomplish it through numerical simulations of all-to-all connected networks. Tiered synchronization paths are observed for certain selections of parameter values. To investigate the mechanism behind the numerical outcomes, we utilize the Ott-Antonsen reduction technique, which reduces the dimension of the entire network. Next, we examine the reduced order model and identify the bifurcations at various parameter regimes that generally guarantee three distinct kinds of synchronization transitions. This analytical procedure can be used to fine-tune the parameters and acquire the necessary transition routes to synchronization. We note that in some parameter regimes, the phase-lag parameter and higher-order coupling strength favor first-order and tiered synchronization techniques.
 
\section{Phase Oscillator Model}
In this paper, we consider an adaptive system of $N$ coupled Sakaguchi-Kuramoto phase oscillators interacting through pairwise and triadic connection. The dynamics of the system is governed by the equations
\begin{eqnarray}
\dot{\theta_i}=\omega_i&+&\frac{K_1 r_1^a}{N} \sum_{j=1}^{N}\sin(\theta_j-\theta_i-\beta)      \nonumber\\ 
&+&\frac{K_2r_1^b}{N^2} \sum_{j=1}^N \sum_{k=1}^N \sin(2\theta_j-\theta_k-\theta_i-\beta),  \label{SKHOI}\\ 
&&\hspace*{3.5cm} i=1,2,\dots,N,  \nonumber
\end{eqnarray}
where $\theta_i$ is the phase and $\omega_i$ is the natural frequency of the $i$th oscillator. $K_1$ and $K_2$ are the coupling strengths corresponding to the pairwise and triadic interactions respectively. $a$ and $b$ are two real constants, and $\beta \in [0,\frac{\pi}{2})$, denotes the phase frustration or phase-lag in the system. The natural frequencies $\omega_i$ are drawn from a distribution $g(\omega)$. The level of synchronization are measured by two global order parameters $r_1$ and $r_2$ corresponding to pairwise and higher-order interactions respectively and are defined by  
\begin{eqnarray}
\label{order_parameter}
z_1=r_1e^{i\psi_1}&=&\frac{1}{N}\sum_{j=1}^{N}e^{i\theta_j},\\
~\mathrm{and}~z_2 &=&r_2e^{i\psi_2}=\frac{1}{N}\sum_{j=1}^{N}e^{2i\theta_j}.
\end{eqnarray}
The complex valued order parameters $z_1$ and $z_2$ describe the macroscopic dynamics of the whole oscillator population. $\psi_1$ and $\psi_2$ are the average phases of the oscillators. Here $r_1=0$ indicates the incoherence state and $r_1=1$ indicates the perfect synchronization state, whereas, $r_2$ measures the 2-cluster synchronization states of the oscillators.

\section{Numerical Observations}\label{NS}
First we numerically simulate the system (\ref{SKHOI}) to understand the effect of the phase-lag $\beta$ on the transition to synchronization in the presence of global order parameter ($r_1$) adaptation . 
The numerical simulation of the system (\ref{SKHOI}) for a network of size $N=1000$ is performed by using the $4$-th order Runge-Kutta method. Intrinsic frequencies of the phase oscillators are drawn from a Lorentzian distribution with mean $0$ and half-width $1$.
\begin{figure}[h!]
\includegraphics[width=9cm]{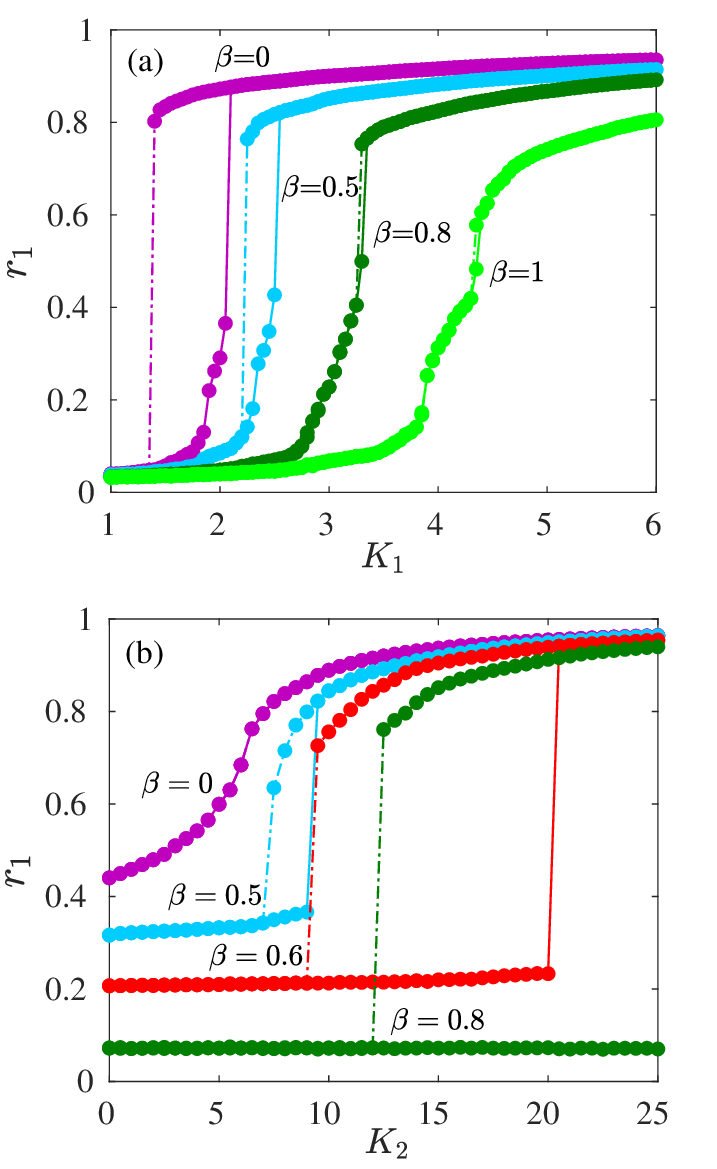}
\caption{Synchronization profile as obtained from the forward (filled circles connected by solid lines) and backward (filled circles connected by dashed dot lines) numerical continuation of the solutions of the system (\ref{SKHOI}) for $a=0$ and $b=2$. (a) Variation of $r_1$ with $K_1$ for $K_2=10$ and different $\beta$. (b) Variation of $r_1$ with $K_2$ for $K_1=2.5$ and different $\beta$. The purple, sky blue, red, green and light green colors respectively represent the curves for $\beta=0$, $0.5$, $0.6$, $0.8$ and $1$.}
\label{fig1}
\end{figure}

For primary investigation, we adapt the order parameter $r_1$ with the higher-order coupling only, i.e. we set $a=0$ and $b \neq 0$ in the equation~(\ref{SKHOI}). The transitions are then studied by looking at the variation of the order parameter $r_1$ either (i) with  the pairwise coupling $K_1$ for fixed $K_2$, $b$ and $\beta$ or (ii) with  the triadic coupling $K_2$ for fixed $K_1$, $b$ and $\beta$. In both the cases, forward and backward continuations of the solutions are done by setting $a=0$ and $b=2$.

In case (i), we take $K_2=10$ and compute the variation of $r_1$ with $K_1$ for different values of $\beta$. The forward continuation for each $\beta$ starts with $K_1=0$ by drawing the initial phases of the oscillators uniformly from the range $-\pi$ to $\pi$. The value of $r_1$ is then computed after removing the transients. Subsequently, the value of $K_1$ is increased in small steps till $K_1 = 6$ and in each step the system (\ref{SKHOI}) is integrated using the last point of previous solution trajectory as the initial condition.   While, for backward continuation, for each $\beta$, simulation starts with $K_1 =6$ and the value $K_1$ is then reduced in small steps, and in each case, the last point on the solution trajectory of the previous solution is used as initial condition for the present simulation. Thus, the variations of $r_1$ with $K_1$ are obtained for different $\beta$ both for the forward and backward continuations.

FIG.~\ref{fig1}(a) shows the variation of $r_1$ with $K_1$ in the range $1$ to $6$ for four different $\beta$, namely, $\beta = 0, 0.5, 0.8$ and $1$. The forward continuation data points are denoted by filled circles of different colors connected by solid lines, while, the backward continuation data points are shown with filled circles connected by dashed dot lines. From the figure it is seen that for all values of $\beta$, the system remains in desynchronized state ($r_1\approx 0$) near $K_1 = 1$. The order parameter follows different paths during the forward and backward continuation of the solutions which results in a hysteresis loop for lower values of $\beta$. However, the width of the hysteresis loop is gradually decreased with the increment of $\beta$ and the transition becomes continuous from discontinuous. It is also observed that for the discontinuous transitions, the forward or backward or both transitions from incoherence to coherence or vice versa occurs via an intermediate weak synchronization state. This type of discontinuous transition involving a continuous path of weekly synchronized states is termed as tiered transition in the literature~\cite{rajwani2023tiered,skardal2022tiered,li2019clustering,li2019synchronization}. As the value of $\beta$ is increased the length of the continuous path expands and persists over a broader range of the coupling strength $K_1$ and as $\beta$ is raised beyond a critical value, continuous transition to synchronization is observed.

On the other hand, in the case (ii), as $K_2$ is varied by fixing $K_1=2.5$, we observe a different scenario. Note that the forward and backward continuation of the solutions are done similarly as those have been done for the case (i). The variation of the global order parameter $r_1$ with $K_2$ for different $\beta$ are shown in the FIG.~\ref{fig1}(b). For $\beta=0$, the forward and backward continuation follow the same path and the transition to synchronization is continuous.  As $\beta$ is increased, the transition becomes discontinuous involving partially synchronized states. For relatively lower values of $\beta$, the discontinuous transition occurs involving tiered synchronization states. However, with the increase of $\beta$, the transition increasingly becomes explosive in nature. 

Thus, from the simulation results discussed above, the observation of the occurrence of tiered synchronization due to the adaptation of order parameter value in the triadic coupling in presence of phase-lag term appears to be quite interesting. To understand  the origin of such dynamical behaviours of the considered system, we proceed with an analytical treatment in the next section. 

\section{Low Dimensional Modelling}
We now use the Ott-Antonsen ansatz~\cite{ott2008low} to reduce the dimension of the globally coupled system of equations (\ref{SKHOI}). To proceed further, we simplify the system (\ref{SKHOI}) and rewrite it in terms of the order parameters defined in (\ref{order_parameter}) as  
\begin{equation}
\label{reduction_model}
\dot{\theta_i}=\omega_i+\frac{1}{2i}\left[e^{-i(\theta_i+\beta)}H-e^{i(\theta_i+\beta)}\bar{H}\right],
\end{equation}
where $H=K_1r_1^a z_1+K_2 r_1^b z_2\bar{z}_1$ and $\bar{H}$ denotes the complex conjugate of $H$. Considering the thermodynamic limit as $N \rightarrow \infty$, let the density of the oscillators at time $t$ with phase $\theta$ and frequency $\omega$ be given by the continuous function $f(\theta,\omega,t)$, and we normalize it as
\begin{equation*}
\int_0^{2\pi} f(\theta,\omega,t) d\theta =g (\omega).
\end{equation*}
Next we take $g(\omega)$ as Lorentzian distribution given by $g(\omega)=\frac{\Delta}{\pi[\Delta^2+(\omega-\omega_0^2)]}$ and expanding $f(\theta,\omega,t)$ in a Fourier series in $\theta$, we obtain
\begin{equation}
f=\frac{g(\omega)}{2\pi}\left \lbrace 1+\sum_{n=1}^{\infty}\left[f_n(\omega,t)e^{in\theta}+\hat{f_n}(\omega,t)e^{-in\theta}\right] \right\rbrace, \label{ott-FS}
\end{equation}
where $f_n(\omega,t)$ is the co-efficient of the $n$-th term of the series and $\hat{f_n}$ stands for the complex conjugate of $f_n(\omega,t)$. Now following~\cite{ott2008low}, we consider an additional ansatz $f_n=\alpha^n (\omega,t)$, where $\alpha$ is an analytic function satisfying the condition $\abs{\alpha}\leq 1$ for convergence.

Since we are dealing with a static network, to maintain the conservation of the oscillators in the network, the density function $f$ satisfies the continuity equation
\begin{equation}
\label{continuty_eq}
\frac{\partial f}{\partial t}+\frac{\partial}{\partial \theta}{(fv)}=0,
\end{equation}
where, $v=\frac{d\theta}{dt}$ is the velocity field on the circle that drives the dynamics of $f$ and is given by the equation~(\ref{reduction_model}). Substituting the expression of $f$ from the equation~(\ref{ott-FS}) into the continuity equation (\ref{continuty_eq}), we obtain the one dimensional differential equation for $\alpha$ as
\begin{equation}
\label{eq_alpha}
\dot{\alpha}+i\alpha\omega+\frac{1}{2}\left[H\alpha^2e^{-i\beta}-\bar{H}e^{i\beta}\right]=0.
\end{equation} 
In the continuum limit, the complex order parameters can be written as 
\begin{eqnarray}
z_1 &=& \int_{-\infty}^{\infty}\int_0^{2\pi}f(\theta,\omega,t)e^{i\theta}d\theta d\omega,  \nonumber \\ 
&=& \int_{-\infty}^{\infty} g(\omega)\bar{\alpha} d\omega.
\end{eqnarray} 
To reduce the dimension of our model, we now introduce one more restriction on our assumed form of $f$; we consider here that $\alpha(\omega,t)$ has no singularities in the lower half of the $\omega$-plane, and that $\abs{\alpha(\omega,t)}\to 0$ as $\rm{Im}(\omega) \to -\infty$. Therefore, the above integration can now be done by taking a close contour in the lower half the $\omega$-plane and for the Lorentzian distribution, we obtain $z_1=\bar{\alpha}(\omega_0-i \Delta, t)$. 

In a similar manner, we can find the order parameter $z_2$
as 
\begin{eqnarray}
z_2 &=& \int_{-\infty}^{\infty}\int_0^{2\pi}f(\theta,\omega,t)e^{2i\theta}d\theta d\omega,  \nonumber \\ 
&=& \int_{-\infty}^{\infty} g(\omega)\bar{\alpha}^2 d\omega, \nonumber \\
&=& \bar{\alpha}^2(\omega_0-i \Delta, t), \nonumber \\
&=& z_1^2.
\end{eqnarray} 
This provides the relation between two order parameters $z_1$ and $z_2$.
Now setting $\omega=\omega_0-i\Delta$ in the equation (\ref{eq_alpha}) we obtain the evolution equation for $z_1$ as
\begin{eqnarray} \label{eq_order_parameter_z}
\dot{z}_1=iz_1\omega_0-z_1\Delta+\frac{1}{2}[(K_1 r_1^a z_1+K_2 r_1^b z_1^2\bar{z}_1)e^{-i\beta} \nonumber \\
-z_1^2(K_1 r_1^a \bar{z}_1+K_2 r_1^b {\bar{z}_1}^2z_1)e^{i\beta}].
\end{eqnarray}
Substituting the relation $z_1=r_1e^{i\psi_1}$ in the equation (\ref{eq_order_parameter_z}) and comparing the real as well as imaginary parts we obtain the $2$-dimensional system of first order ordinary nonlinear differential equations
\begin{eqnarray}
\dot{r}_1&=& -\Delta r_1+\frac{\cos\beta}{2}(K_1 r_1^{a+1}+K_2 r_1^{b+3})[1- r_1^2], \label{eq_r} \\
\dot{\psi_1}&=& \omega_0-\frac{\sin\beta}{2}(K_1 r_1^a+K_2r_1^{b+2})[ 1+r_1^2],  
\label{eq_psi}
\end{eqnarray}
to investigate the dynamics of the system (\ref{SKHOI}). Subsequently we use the reduced order model (ROM) given by the equations (\ref{eq_r})-(\ref{eq_psi}) for a detailed analysis. 

\section{ROM Vs numerical simulation}
To understand the origin of different kinds of transitions to synchronization including the ones observed in the numerical simulation, described in the section (\ref{NS}) we perform detailed bifurcation analysis of the reduced order model (\ref{eq_r}) using the MATCONT software~\cite{dhooge2003matcont}. The results are then compared with that of the numerical simulations of the full system (\ref{SKHOI}) at different points of the parameter space for validation. 

\begin{figure}[h!]
\includegraphics[width=9cm]{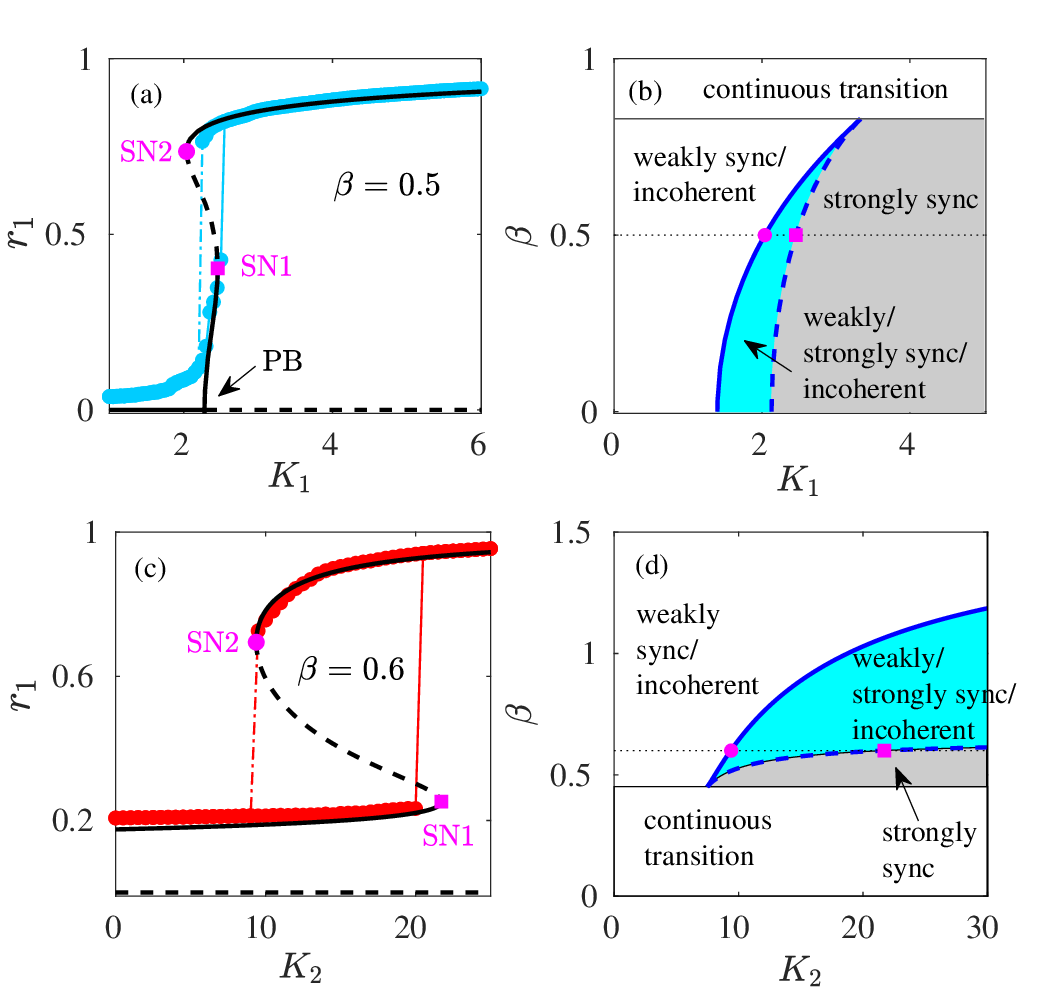}
\caption{Bifurcation diagrams constructed from the ROM (\ref{eq_r}) for $a=0$ and $b=2$: (a) $r_1$ vs $K_1$ for $K_2=10$ and $\beta=0.5$, and (c) $r_1$ vs $K_2$ for $K_1=2.5$ and $\beta=0.6$. The stable and unstable branches are shown with solid and dashed black curves respectively. The corresponding data points obtained from the numerical simulation of (\ref{SKHOI}) are shown with filled circles connected by solid (forward) and dashed dot (backward) lines. The saddle-node and pitchfork bifurcation points on the bifurcation curve are marked with SN1, SN2 and PB respectively.  Two parameter diagrams showing different synchronization regimes on the: (b) $K_1 -\beta$ and (d) $K_2 -\beta$ planes. The solid (dashed) blue curve indicates the backward (forward) saddle node points for different $\beta$. The pink squares and dots respectively indicate the forward and backward saddle node points shown in the bifurcation diagrams (a) and (c). }
\label{fig3}
\end{figure}

Before proceeding further, we focus on the synchronization curves obtained from the numerical simulation of the system (\ref{SKHOI}) and presented in the FIG.~\ref{fig1}(a) for $\beta = 0.5$ (sky blue), and figure~\ref{fig1}(b) for $0.6$ (green) respectively where transitions are related with tiered synchronization. To understand the origin of such interesting transition in the presence of triadic interactions, we construct two bifurcation diagrams using the ROM (\ref{eq_r}) for the same set of parameter values used in the numerical simulation. The bifurcation diagrams are shown in the figures~\ref{fig3}(a) and (c) respectively along with the corresponding numerical simulation data. 

In the FIG.~\ref{fig3}(a), the solid and dashed black curves respectively represent the stable and unstable solutions as obtained from the ROM (\ref{eq_r}). The $r_1 = 0$ solution is stable for lower values of $K_1$ and it becomes unstable through supercritical pitchfork bifurcation (PB) at $K_1 = 2.279$. A stable branch (solid black curve) with nonzero $r_1$ is generated from there and the zero solution continue to exist as unstable one for $K_1 > 2.279$. The nonzero branch move forward a bit to undergo a saddle-node (SN1) bifurcation at $K_1 = 2.457$ (pink square). The resulting unstable blanch (dashed black curve) move backward and undergoes another saddle-node (SN2) bifurcation (pink dot) at $K_1 = 2.038$.  The stable branch originated out of the SN2 bifurcation then continue to exist for higher values of $K_1$. Now the occurrences of a pair of saddle-node bifurcations leads to discontinuous transition to synchronization in the system with the variation of $K_1$ in the presence of adaptation in the triadic interaction. 

For the validation of the ROM results we now superpose the data obtained from the numerical simulation both for forward and backward continuation of the solutions on top of the bifurcation diagram. The forward and backward continuation data points are presented with cyan dots connected by solid and dashed dot lines respectively. The numerical simulation data points show a very close match with the ROM results.  We note that between the SN2 and SN1 points, three stable states of the system, namely, incoherent, weakly synchronized and strongly synchronized coexist. As a result, with the variation of $K_1$, transition occurs in the system through tiered 
synchronization states. Thus, we claim that the occurrence of a pair of saddle-node bifurcations is responsible for the appearance of discontinuous transition in the system involving tiered synchronization.
\begin{figure}[h!]
\includegraphics[width=9cm]{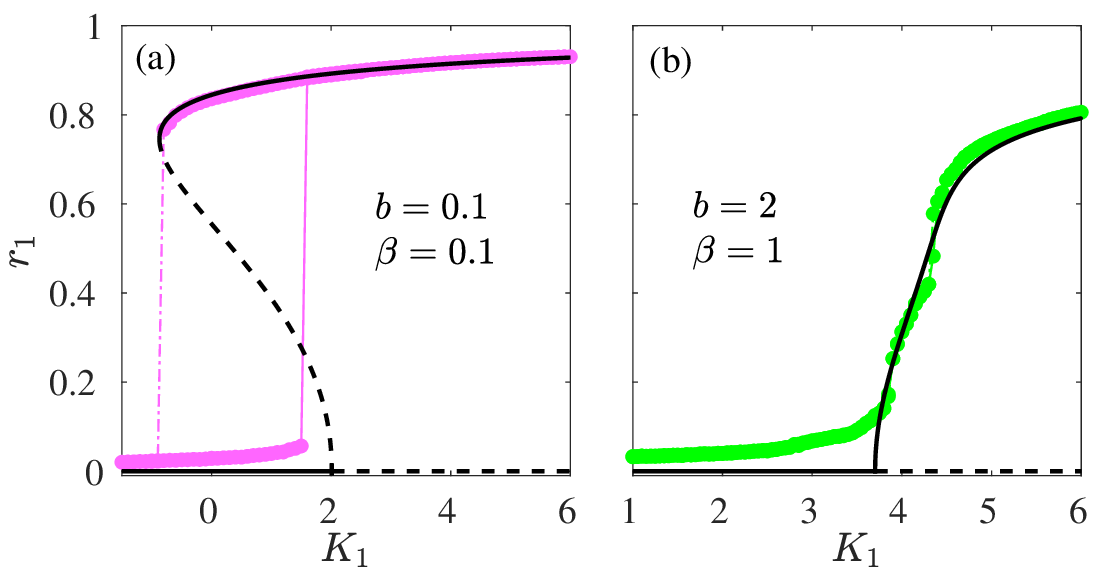}
\caption{Bifurcation diagram constructed from the ROM: $r_1$ vs $K_1$ for (a) $b=0.1$, $\beta=0.1$ and (b) $b=2$, $\beta=1$ with fixed $K_2=10$, $a=0$. The solid and dashed black curves respectively show the stable and unstable fixed points of the ROM (\ref{eq_r}). The data points obtained from the numerical simulation of the system (\ref{SKHOI}) are shown with filled colored circles.} 
\label{fig3_1}
\end{figure}

For detailed understanding, we now prepare a two parameter diagram using the ROM showing different synchronization regimes on the $K_1-\beta$ plane (see FIG.~\ref{fig3}(b)). The dashed and solid blue curves in the diagram respectively show the locations of the successive saddle-node bifurcation points as a function of $\beta$. The pink square and dot on these lines correspond to the SN1 and SN2 points of the FIG.~\ref{fig3}(a) respectively. As long as two distinct saddle-node bifurcation points exist, the transition becomes discontinuous involving tiered synchronization. The saddle-node curves meet at $\beta = 0.8298$, and the transition becomes continuous thereafter. The different synchronization regimes are clearly demarcated and shown in the FIG.~\ref{fig3}(b). Thus, from the two parameter diagram we observe that the phase-lag parameter $\beta$ inhibits discontinuous transition for fixed $K_2$. This gradual suppression of the discontinuous transition by $\beta$ is consistent with the numerical observations presented in the section (\ref{NS}). 

Next, the variation of the order parameter $r_1$ for different stable and unstable solutions with $K_2$ for fixed $\beta = 0.6$ and $K_1 = 2.5$ are shown in the bifurcation diagram presented in the FIG.~\ref{fig3}(c).  The stable and unstable solutions are shown with solid and dashed black curves. The $r_1 = 0$ solution is unstable in the entire range of $K_2.$ The non-zero solutions ($r_1 \neq 0$) show a pair of saddle-node bifurcations marked by pink square and dot. In between these two saddle-node bifurcation points, the system is bistable which introduces a large hysteresis loop leading to discontinuous transition. Here also the transition involves tiered synchronization. The data points obtained from the numerical simulations of the full system (\ref{SKHOI}) are then superposed on the bifurcation diagram. The forward and backward continuation data are shown with red dots connected by solid and dashed dot red lines respectively. In this case also the numerical simulation results show a very close match with the ROM results. 

The detailed bifurcation scenario can be seen from the two parameter diagram shown in the FIG.~\ref{fig3}(d) presenting different synchronization regimes on the $K_2 - \beta$ plane for fixed $K_1 = 2.5$, $a=0$ and $b =2$. The solid and dashed blue curved respectively show the variation of the backward and forward saddle-node bifurcation points with $\beta$. The pink square and dot correspond to the saddle-node bifurcation points shown in the FIG.~\ref{fig3}(c). We note here that unlike the two parameter diagram shown in the FIG~.~\ref{fig3}(b), the region enclosed by the blue curves is not closed. Both the forward and backward saddle-node bifurcation curves diverge towards very high value of $K_2$ with the enhancement of $\beta$, where the rate of divergence of the forward SN point is much faster compared to the backward one. This is also consistent with the numerical observation shown in the FIG.~\ref{fig1}(b). 

\begin{figure}[h!]
\includegraphics[width=9cm]{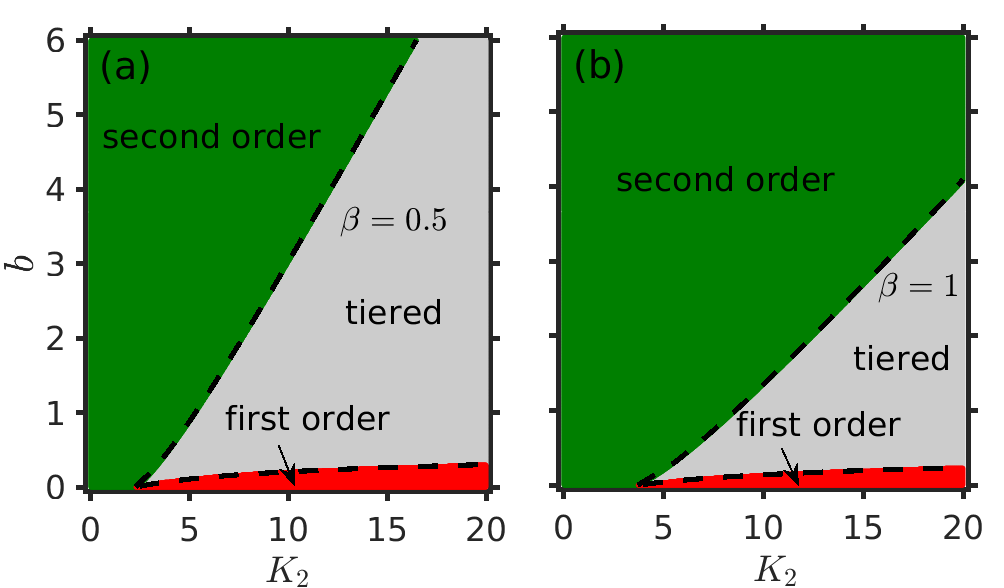}
\caption{Different synchronization transition regimes on the $K_2-b$ plane obtained with the variation of $K_1$ in the ROM for (a) $\beta=0.5$ and (b) $\beta=1$ in the absence of adaptation in the pairwise coupling ($a=0$). First order (red), tiered (grey) and second order (green) regimes are separated by dashed black curves, indicates the critical parameter sets.} 
\label{lag_vs_b}
\end{figure}

In the FIG.\ref{fig3} (a) and (c), we have either varied the parameters $K_1$ or $K_2$ keeping the other parameters fixed and observed tiered synchronization. From the bifurcation point of view, apart from tiered synchronization, we have also have identified the points in the parameter space where we observe second order and explosive transitions to synchronization both in the ROM and numerical simulations. FIG.~\ref{fig3_1}(a) and (b) show the bifurcation diagrams obtained from the ROM (\ref{eq_r}) where explosive and second order continuous transitions are observed with the variation of $K_1$. The numerically obtained data points for the same set of parameter values are also plotted in these figures which show a very close match. Inspired by this observation, we now wish to develop a deeper understanding of the parameter space using the ROM. In this context, a closer look at the bifurcation curves presented in the FIGs.~\ref{fig3}(a),~\ref{fig3_1}(a) and (b), reveal that for tiered, explosive and continuous transitions to synchronization, the synchronization curves possess three, two and one extrema points respectively of the graph of $K_1$ as a function of $r_1$ given by 
\begin{eqnarray}
    K_1=\frac{2}{r_1^a(1-r_1^2)\cos \beta}-K_2 r_1^{(b+2-a)}.
    \label{eq_K1}
\end{eqnarray}
Similarly, as $K_2$ is varied by keeping the other parameters fixed, the number of extrema of the function  
\begin{eqnarray}
    K_2=\frac{2}{r_1^{(b+2)}(1-r_1^2)\cos \beta}-\frac{K_1r_1^a}{r_1^{(b+2)}}
    \label{eq_K2}
\end{eqnarray}
will determine the nature of transitions in the system. Thus, using the equations (\ref{eq_K1}) and (\ref{eq_K2}) we prepare the diagrams shown in the figures (\ref{lag_vs_b}), (\ref{lag_vs_b1}) and (\ref{b_vs_K1}. Different transition regions are nicely depicted in these figures. 

FIG.~\ref{lag_vs_b} shows the transition regimes on the $K_2-b$ plane for $\beta = 0.5$ and $1$ with the variation of $K_1$.  

\begin{figure}[h!]
\includegraphics[width=9cm]{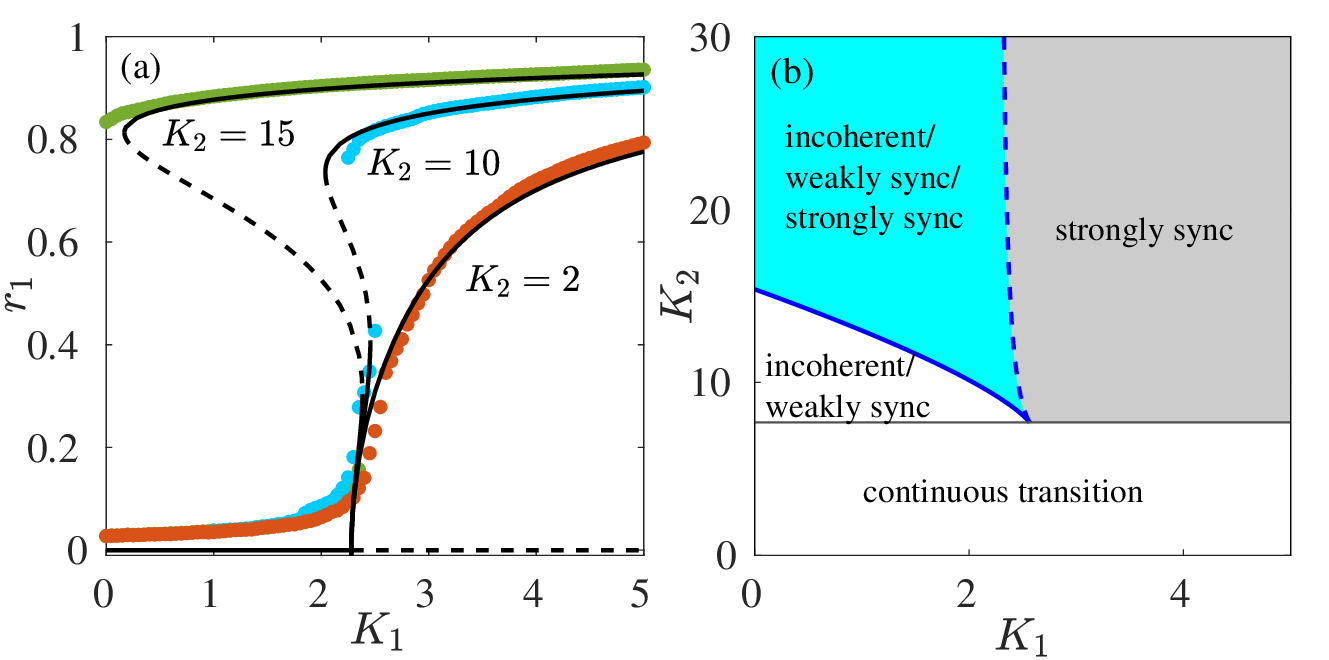}
\caption{Bifurcation diagram constructed from the ROM: (a) Synchronization profile for different $K_2$ with fixed $a=0$, $b=2$ and $\beta=0.5$. Numerically calculated points (colored filled circles) are plotted along with the solid black curves obtained from the ROM. Dashed black curve indicates the unstable states. (b) Two parameter diagram showing different synchronization regimes on the $K_1 - K_2$ plane. The existence of both forward (dashed blue curve) and backward (solid blue curve) saddle node points implies tiered synchronization states in the system.}
\label{K1_vs_r_diff_K2}
\end{figure}

The figure clearly shows that $\beta$ promotes second order transition in the absence of adaptation in the pairwise coupling when $K_1$ is varied for fixed $K_2$ and $b$. On the other hand, $\beta$ suppresses the tiered and first order transitions for similar variations of the parameters. To explore the transitions in more detail, we take the point $(K_2, b) = (2,2)$ inside the green region of the FIG.~\ref{lag_vs_b} where second order transition occurs with the variation of $K_1$ and then we move horizontally to enter the tiered region.  The bifurcation diagrams obtained from the ROM (\ref{eq_r}) alongside the numerical simulation data are shown in the FIG.~\ref{K1_vs_r_diff_K2} (a) for three values of $K_2$ for fixed $b = 2$. A very good match of the bifurcation structure with the numerical simulation data is observed. As mentioned earlier, the tiered synchronization are characterized by a pair of saddle-node bifurcations. The variation of those saddle-node bifurcation points on the $K_1 - K_2$ planes are shown with the solid and dashed blue lines in the FIG.~\ref{K1_vs_r_diff_K2}(b). These lines separate different synchronization regimes on that plane.  

\begin{figure}[h!]
\includegraphics[width=9cm]{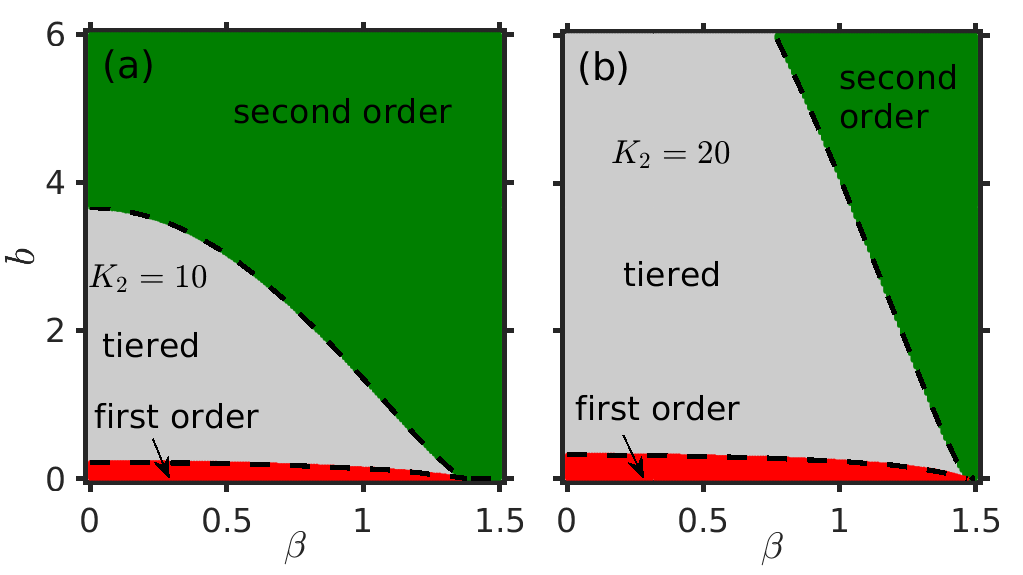}
\caption{Different synchronization transition regimes on the $\beta-b$ plane obtained with the variation of $K_1$ in the ROM for (a) $K_2=10$ and (b) $K_2=20$ in the absence of adaptation in the pairwise coupling ($a=0$). First order (red), tiered (grey) and second order (green) regimes are separated by dashed black curves, indicates the critical parameter sets.}

\label{lag_vs_b1}
\end{figure}

\begin{figure}[h!]
\includegraphics[width=9cm]{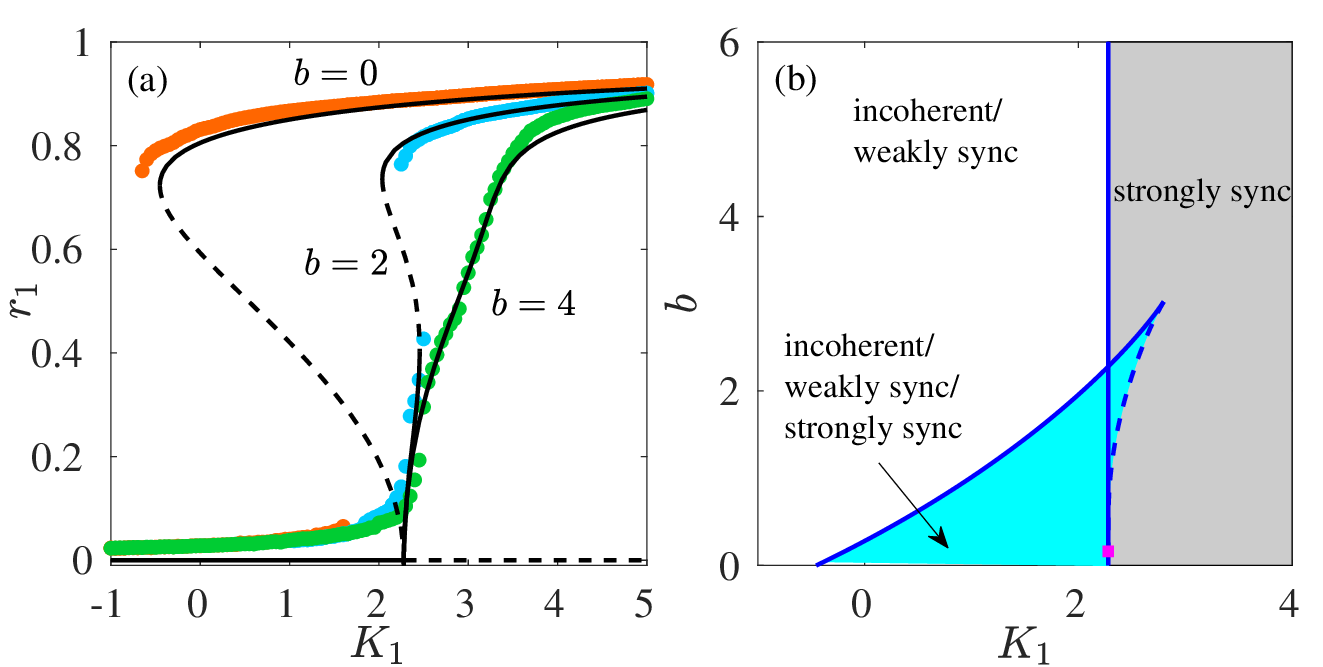}
\caption{Bifurcation diagram constructed from the ROM: (a) $r_1$ vs $K_1$ for different $b$ with $K_2=10$, $a=0$ and $\beta=0.5$. The solid and dashed black curves respectively show the stable and unstable fixed points of the ROM (\ref{eq_r}). The data points obtained from the numerical simulation of the system (\ref{SKHOI}) are shown with filled circles of different colors. (b) Two parameter diagram showing different synchronization regimes on the $K_1-b$ plane. Solid and dashed blue line respectively show the backward and forward saddle node points.}
\label{K1_vs_r_diff_b}
\end{figure}

FIG.~\ref{lag_vs_b1} determined from the equation (\ref{eq_K1}) shows the synchronization transition regimes on the $\beta - b$ plane for two values of $K_2$. In this case, $K_2$ suppresses the second order transition, and promotes first order and tiered transitions. To compare the ROM results with that of the numerical simulations, we take three points $(0.5, 0)$, $(0.5, 2)$ and $(0.5,4)$ from the $\beta-b$ plane shown in the FIG.~\ref{lag_vs_b1}(a) where, respectively, first order, tiered and second order transitions to synchronization occur in the ROM. We then construct bifurcation diagrams from the ROM using the parameter sets corresponding to these points and show in the FIG.~\ref{K1_vs_r_diff_b}(a) alongside the numerical simulation data points. A very close match between the ROM and full system results are observed. We then go ahead to identify different synchronization regimes on the $K_1-b$ plane using the ROM (\ref{eq_r}) and present in the FIG.~\ref{K1_vs_r_diff_b}(b). The solid and dashed blue curves enclosing the cyan region in the figure shows the backward and forward saddle-node points associated with the tiered transition respectively and the blue solid line denotes the location of pitchfork bifurcation points. Other synchronization regimes are also clearly depicted in this figure.
\begin{figure}[h!]
\includegraphics[width=9cm]{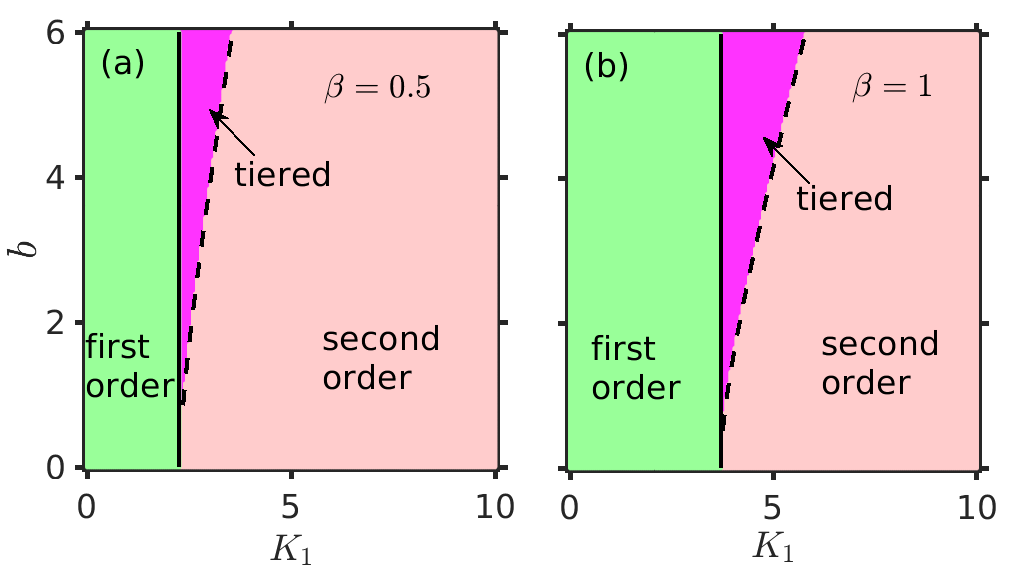}
\caption{Different synchronization transition regimes on the $K_1-b$ plane obtained with the variation of $K_2$ in the ROM for (a) $\beta=0.5$ and (b) $\beta=1$ setting $a=0$. The solid and dashed black curves separate the regimes of first order (light green), tiered (magenta) and second order (light red) synchronization transitions.}
\label{b_vs_K1}
\end{figure}

\begin{figure}[h!]
\includegraphics[width=9cm]{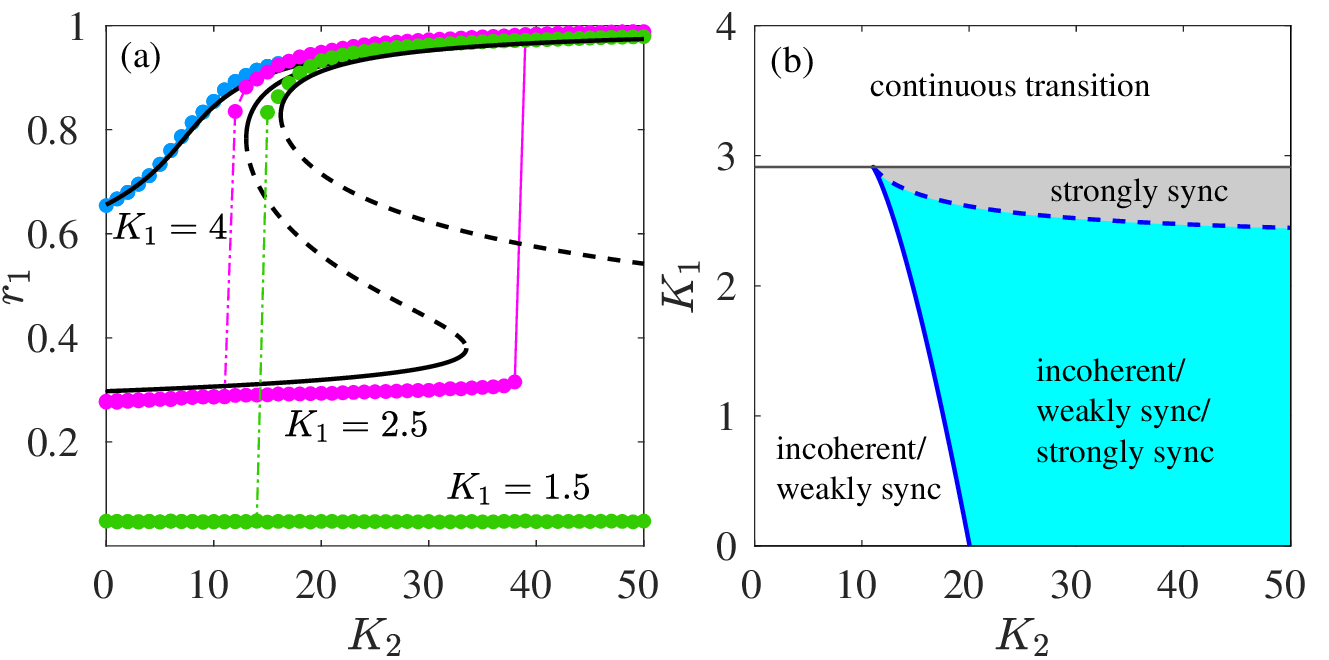}
\caption{(a) Bifurcation diagrams constructed using the ROM for $a=0$, $b=3.5$ and $\beta=0.5$ and different values of $K_1$. The solid and dashed black curves respectively represent the stable and unstable fixed points of the ROM.  The data points obtained by the numerical simulation of the system (\ref{SKHOI}) are also shown on top of the bifurcation curves. (b) Two parameter diagram showing different synchronization regimes as obtained from the ROM on the $K_2-K_1$ plane. Solid and dashed blue curves respectively show the backward and forward saddle node bifurcation points of the ROM.}
\label{K2_vs_r_diff_K1}
\end{figure}

Next we use the equation (\ref{eq_K2}) to investigate the effect of phase frustration on the transitions to synchronization. FIG.~\ref{b_vs_K1} shows the two parameter diagram prepared using the equation (\ref{eq_K2}) depicting various transition regimes on the $K_1 - b$ plane for $\beta = 0.5$ and $1$ with the variation of $K_2$. From the diagram it is seen that in this case, phase-lag promotes first order as well as tiered transition to synchronization, while, it suppresses second order transition. In this case also, we take three points, namely, $(1.5,3.5)$, $(2.5,3.5)$ and $(4,3.5)$ from the $K_1 - b$ plane shown in the FIG.~\ref{b_vs_K1}(a) where first order, tiered and second order transitions respectively are observed in the ROM. We construct bifurcation diagrams from the ROM using the parameter set corresponding to those points and present in the FIG.~\ref{K2_vs_r_diff_K1}(a) along with the associated numerical simulation data points obtained from the full system (\ref{SKHOI}). The comparison of the ROM results with that of the numerical simulation of the full system not only show similar transition type but also show a very close match in the order parameter values. As done in the previous two cases, here also we identify different synchronization regimes on the $K_2 - K_1$ plane using the ROM and present in the FIG.~\ref{K2_vs_r_diff_K1}(b). 

\begin{figure}[h!]
\includegraphics[width=9cm]{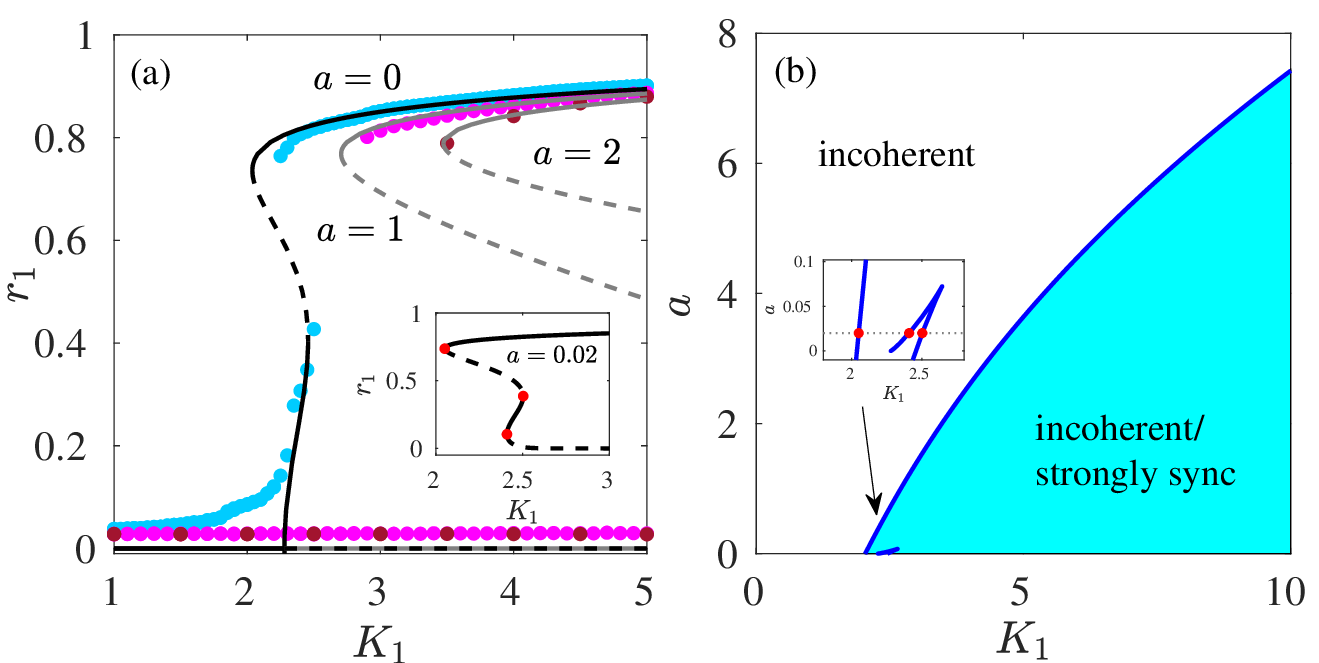}
\caption{(a) Bifurcation diagram constructed from the ROM for $K_2=10$, $b=2$ and $\beta=0.5$ and different values of $a$. The stable and unstable fixed points of the ROM are shown with solid and dashed black curves. The corresponding $r_1$ values obtained from numerical simulation of the system (\ref{SKHOI}) are shown with filled circles of different colors. The inset presents the bifurcation diagram for $a=0.02$ showing three saddle node bifurcation points with red dots.  (b) Two parameter diagram depicting different synchronization regimes on the $K_1-a$ plane. The inset shows an enlarged view of the indicated region. Three red dots correspond to the saddle-node points shown in the inset of (a). } 
\label{K1_vs_r_diff_a}
\end{figure}
So far we have investigated different types of transitions to various synchronization states in the absence of the adaptation in the pairwise coupling term ($a=0$). We have extensively used the ROM (\ref{eq_r}) for the investigation and validated the results with the ones obtained from the numerical simulation of the full system (\ref{SKHOI}). The analysis revealed a clear picture of the transition scenario involving continuous as well as discontinuous transitions in a wide region of the parameter space. The discontinuous transitions include both explosive and tiered synchronizations. 

Encouraged by the satisfactory match of the results, we now consider the case where the adaptation of the order parameter is present in both pairwise and higher-order term. The FIG.~\ref{K1_vs_r_diff_a} demonstrates the role of the adaptation exponent $a$ on the transition to synchronization. In the FIG.~\ref{K1_vs_r_diff_a}(a), the bifurcation diagrams prepared from the ROM are shown for different values of $a$. The values of the other parameters $K2$, $b$ and $\beta$ are kept fixed. For this parameter choice, the system shows tiered transition for $a=0$ and explosive transition for other values of $a$. The numerically computed order parameters from the full system (\ref{SKHOI}) are seen to fit nicely with the ROM curve. Beside this synchronization diagram, we also plot the stability diagram in Fig.\ref{K1_vs_r_diff_a}(b) to look into the bifurcation points. In the figure, the solid blue curve represents the saddle-node bifurcation points as a function of $a$. For very small values of $a$ ($0<a\leq 0.07238$), more than one saddle-node bifurcation points are obtained and for $a > 0.07238$, there is only one SN point. Thus, for $0< a \leq0.07238$, tiered synchronization is observed, while, for $a > 0.07238$, we obtain explosive synchronization only. The blue curve in the figure clearly divides the region into different synchronization regimes. 

The exhibition of different types of transitions to synchronization in the system depends on the complex interaction among the pairwise and triadic coupling terms, and the phase frustration parameter $\beta$. The phase frustration  $\beta$ and the pairwise coupling strength $K_1$ are known to promote second order transition by inhibiting discontinuous transition~\cite{skardal2020higher}, while, the higher-order coupling $K_2$ promotes discontinuous transition in the system~\cite{dutta2023impact}. 
Since the global order parameter $r_1$ ranges from $0$ to $1$, for any value of $ a,b \geq 0$, $0 \leq r_1^a, r_1^b \leq 1$. Multiplication of $K_1$ by $r_1^a$ and $K_2$ by $r_1^b$ reduces the respective effective coupling strengths for $a,b > 1$. Thus, relative dominance of the pairwise and triadic coupling terms in the presence of phase frustration depends on the parameter ranges and the type of the transitions are determined accordingly in the system.

\begin{figure}[h!]
\includegraphics[width=9cm]{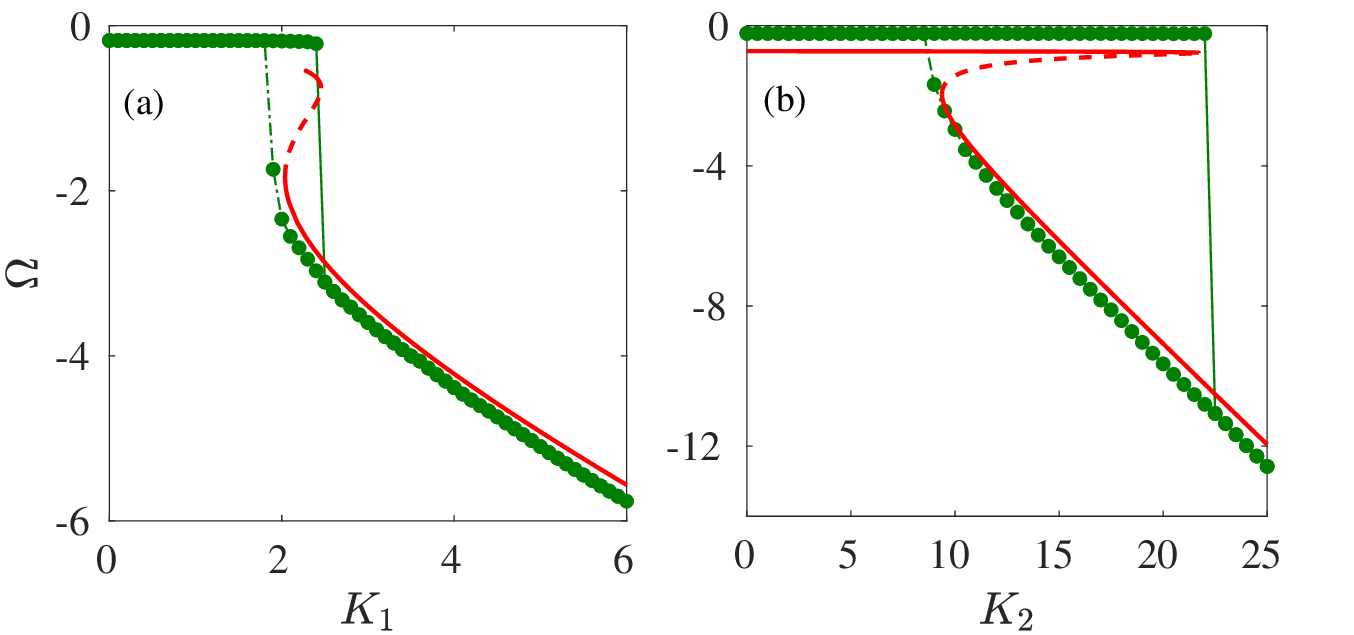}
\caption{(a) Mean frequency $\Omega$ as a function of pairwise coupling strength $K_1$ for phase-lag $\beta=0.5$ and (b) $\Omega$ as a function of higher-order coupling strength $K_2$ for $\beta=0.6$. Filled green circles indicate numerically calculated points and the solid (dashed) red curves indicate the stable (unstable) points calculated from the ROM.}
\label{fig7}
\end{figure}

Finally, we compute the mean field frequency $\Omega$ using the equations (\ref{eq_r}) and (\ref{eq_psi}), and the full system (\ref{SKHOI}). FIG.~(\ref{fig7}) represents the variation of $\Omega$ with both pairwise coupling $K_1$ and higher-order coupling $K_2$. The figure shows that the mean frequency of the system is also closely captured by the reduced order model derived from the full system. 
\section{Conclusions}
In this paper, we have investigated the phenomenon of transition to synchronization in globally coupled system of Sakaguchi-Kuramoto oscillators in the presence of higher-order interactions (up to triangular) and order parameter adaptation. The system involves five control parameters, namely, $K_1$, $K_2$, $a$, $b$ and $\beta$ representing the coupling constants and adaptation exponents in the pairwise and triadic coupling terms, and phase frustration respectively.  

Primary numerical simulations of the full system at some specific points in the parameter space show interesting continuous as well as discontinuous transitions of different types. However, investigating the role of different parameters in determining the type transitions to synchronization in the system by performing numerical simulations only is a challenging task and computationally expensive. Thus, along with the performance of numerical simulations of the full system, we also derive a simple reduced order model of the system using the Ott-Antonsen ansatz to understand the transition scenario in detail in a large region of the parameter space. 

The comprehensive bifurcation analysis of the ROM clearly demarcates different regions of the parameters space exhibiting different types of transitions to synchronization. The analysis reveals a complex dependence of the transitions on the parameters. It is observed that the second order continuous transition is connected with a supercritical pitchfork bifurcation of the ROM. While, the discontinuous teired transition is associated with multiple SN bifurcations along with a supercritical PB and the  first order explosive transition involves a single subcrtical PB alongside a SN bifurcation.  The results obtained from the ROM are compared with that of the numerical simulations at several points of the parameter space and are found to match closely. 

The investigation of the system both numerically as well as analytically shows that the joint role of different parameters of the system in the transition to synchronization can not be understood simply by superposing their individual roles. For example, the phase frustration is known to promote second order transition in the the absence of adaptation and higher order coupling~\cite{sakaguchi1986soluble,kundu2017transition}. However, as seen in the FIG.~\ref{b_vs_K1}, under certain conditions, the phase frustration may promote discontinuous transitions. Nonetheless, in some situations as depicted in the FIG.~\ref{lag_vs_b}, the phase frustration can suppress the discontinuous  transitions and exhibits its usual role in promoting second order transition in the system. Therefore, a general conclusion about the role of the individual parameters on the transitions is found to be difficult. Best way to look at the projections of the parameter space on two dimensional planes as has been done in this paper for a better understanding of the transitions in the parameter space. 

The results of the investigation presented in the paper show that in the sole presence of adaptation in the triadic coupling, as $K_1$ is varied for fixed $b$, $K_2$ and $\beta$, the system exhibits all three types of transitions depending on the values of the fixed parameters as seen in the FIG. \ref{lag_vs_b}(a). It is also seen that with the enhancement of $\beta$, the discontinuous transition regions are suppressed (see FIG. \ref{lag_vs_b}(b)). On the contrary, FIG. \ref{b_vs_K1} suggests the promotion of discontinuous transition by $\beta$ on the $K_1 - b$ plane. Similarly, the promotion of discontinuous transitions by the higher order coupling parameter $K_2$ is evident from the FIG.\ref{K1_vs_r_diff_K2}. Thus, the present analysis provide an elegant methodology to unfold the transition scenario in the system. 

\section*{ACKNOWLEDGEMENTS}
S.D. acknowledges the support from DST, India under the INSPIRE program (Code No. IF190605).

\bibliographystyle{apsrev4-1}
%

\end{document}